\documentclass[a4paper,11pt]{article}
\usepackage{pos}

\usepackage{subcaption}
\usepackage{color,soul}
\usepackage{float}

\usepackage{braket}
\usepackage{blkarray}
\usepackage{url}
\usepackage{adjustbox}

\newcommand{\newsection}[1]{\vspace{-8 pt}\section{#1}\vspace{-8 pt}}

\newcommand{\qhat}{\hat{q}}
\newcommand{\phat}{\hat{p}}

\newcommand{\bhat}{\hat{b}}
\newcommand{\bhatdag}{\hat{b}^\dagger}

\title{Towards quantum simulation of lower-dimensional supersymmetric lattice models}

\author*{Emanuele Mendicelli}
\author{David Schaich}

\affiliation{Department of Mathematical Sciences, University of Liverpool,
Liverpool L69 7ZL, United Kingdom}

\emailAdd{e.mendicelli@liverpool.ac.uk}
\emailAdd{david.schaich@liverpool.ac.uk}

\abstract{Supersymmetric models are grounded in the intriguing concept of a hypothetical symmetry that relates bosonic and fermionic particles. This symmetry has profound implications, offering valuable extensions to the Standard Model of particle physics and fostering connections to theories of quantum gravity. However, lattice studies exploring the non-perturbative features of these models, such as spontaneous supersymmetry breaking and real-time evolution encounter significant challenges, particularly due to the infamous sign problem.

The sign problem obstructs simulations on classical computers, especially when dealing with high-dimensional lattice systems. While one potential solution is to adopt the Hamiltonian formalism, this approach necessitates an exponential increase in classical resources with the number of lattice sites and degrees of freedom, rendering it impractical for large systems. In contrast, quantum hardware offers a promising alternative, as it requires in principle a polynomial amount of resources, making the study of these models more accessible.

In this context, we explore the encoding of lower-dimensional supersymmetric quantum mechanics onto qubits.
We also highlight our ongoing efforts to implement and check the model supersymmetry breaking on an IBM gate-based quantum simulator with and without shot noise, addressing the technical challenges we face and the potential implications of our findings for advancing our understanding of supersymmetry.}

\FullConference{The 41st International Symposium on Lattice Field Theory (LATTICE2024)\\
 28 July - 3 August 2024\\
Liverpool, UK\\}

\begin{document}
\maketitle

\newsection{Introduction and motivations}
Supersymmetric models are built assuming that nature can be symmetric under the exchange of fermions and bosons.  This fruitful idea has given rise to extensions of the Standard Model, established connections with the theory of quantum gravity and overall deepened our understanding of quantum field theory. Lattice studies of supersymmetric models have been successful to characterize non-perturbative aspects \citep{Kadoh:2016eju, Bergner:2016sbv, Schaich:2022xgy}, but the sign-problem obstructs the study of supersymmetry breaking and real-time evolution  \citep{Bergner:2016sbv, Schaich:2022xgy}.  The sign-problem is not present in the Hamiltonian formalism, where classical computers have difficulties to deal with the Hilbert space that grows exponentially. Quantum computers can encode the Hilbert space of a supersymmetric model using in principle a polynomial number of quantum resources, therefore making their usage extremely beneficial.  At the same time, supersymmetric models offer an interesting and challenging testbed for current NISQ era \citep{Preskill:2018jim} hardware and algorithms, because they require the encoding on the hardware of bosonic and fermionic degrees of freedom.
This interplay between quantum computing and supersymmetry can be concretely realized by using the variational quantum eigensolver (VQE) \citep{Peruzzo:2013bzg} algorithm to measure the model's ground state energy, which is a direct check of spontaneous supersymmetry breaking. In recent years, several studies have begun to explore this interplay \citep{Culver:2021rxo, Cai:2022yup, Culver:2023iif, Schaich:2024bmg}, however, \citep{Culver:2021rxo, Culver:2023iif, Schaich:2024bmg} did not account for shot noise.

In this proceeding we continue previous efforts \citep{Culver:2021rxo} in studying supersymmetry breaking of $0+1$ dimensional supersymmetric quantum mechanics (SQM), by presenting results of quantum simulations with and without the presence of shot noise. We start by presenting the SQM model and its qubitization in Section~\ref{SQM_and_qubitization}. Then in Section~\ref{sec:vqe_boxplot} the VQE algorithm is described together with the way of presenting the results. The preliminary results are presented in Section~\ref{sec:vqe_results} followed by a discussion and future prospects in Section~\ref{sec:conclusions}.

\newsection{Supersymmetric quantum mechanics}\label{SQM_and_qubitization}
We consider SQM, which is a discrete Hamiltonian model with one fermion and one boson interacting at a single site in continuous time:
\begin{equation} \label{eq:H_SQM}
H = \frac{1}{2}\left( \phat^2 + [W'(\qhat)]^2 - W''(\qhat) \left[\bhatdag, \bhat\right]\right) 
\end{equation}
where $\phat$ and $\qhat$ are the momentum and position coordinate of the boson, with standard relations $[\qhat,\phat]=i$.
The boson-fermion interaction is expressed by the superpotential $W(\qhat)$, where the prime symbol denotes derivatives with respect to $\qhat$. The 
$\bhat$ and $\bhatdag$ are the fermion annihilation and creation operators, with the standard relation $\lbrace \bhat, \bhatdag \rbrace =1$ and their action on the fermion states is:
\begin{align}
  \bhat \vert 1 \rangle & = \vert 0 \rangle 	& \bhatdag \vert 1 \rangle  & = 0 \cr
  \bhat \vert 0 \rangle & = 0     				& \bhatdag \vert 0 \rangle & = \vert 1 \rangle
\end{align}
The superpotential plays an important role in the model, because its specific shape determines spontaneous supersymmetry breaking or preservation \cite{Cooper:1994eh}. The ones considered here are the Harmonic Oscillator (HO), the Double Well (DW) and the Anharmonic Oscillator (AHO):
\begin{center}
\begin{tabular}{ccc}
Harmonic Oscillator & Anharmonic Oscillator & Double Well \\[5pt]
$W(\qhat)=\frac{1}{2}m\qhat^2$ & $W(\qhat)=\frac{1}{2}m\qhat^2 + \frac{1}{4} g \qhat^4$ & $W(\qhat)=\frac{1}{2}m\qhat^2 +g \left( \frac{1}{3}\qhat^3 + \mu^2 \qhat  \right)$\\[5pt]
[Supersymmetric] & [Supersymmetric] & [Spontaneously breaking] \\
\end{tabular}
\end{center}
where $m$ is the boson mass and by supersymmetry the mass of the fermion; while $g$ and $\mu$ the strength of interactions. These are all made dimensionless using powers of the lattice spacing and set to $m=g=\mu=1$ for simplicity. The supersymmetry preservation or breaking of the model with a given superpotential, can be obtained by calculating the model's ground state energy (i.e. zero preserved, non-zero breaking). This seemingly straightforward task becomes numerically challenging for systems with many states, as it requires an exponentially increasing amount of classical resources to encode the Hamiltonian. In this respect, quantum computers can be advantageous, potentially reducing the problem to polynomial complexity.

In concluding this section, to make evident the supersymmetric nature of the model, we need to introduce the supercharges that are responsible to establish the symmetry between bosonic and fermionic operators:
\begin{align}
Q & = \bhat \left( i \phat + W'(\qhat) \right)  & Q^{\dagger} & = \bhatdag \left( -i \phat + W'(\qhat) \right) & Q^2 = Q^{{\dagger} 2} =0 \,.
\end{align}
The charges can be used to express the Hamiltonian as $2 H=\lbrace Q, Q^{\dagger}  \rbrace$. This characteristic supersymmetric relation has fundamental implication on the energy spectrum. All eigenstates have non-negative energy, $E_\Psi = \langle \Psi \vert H \vert \Psi \rangle =\frac{1}{2}\left( \vert Q \vert \Psi   \rangle \vert ^2  + \vert Q^{\dagger} \vert \Psi \rangle \vert ^2\right)  \geq 0$. 
An eigenstate has $E_\Psi=0$ only if it is annihilated by both supercharges, which means that the symmetry leaves the vacuum invariant, and therefore the system preserves supersymmetry. All the eigenstates with positive energy $E_\psi>0$ come in degenerate pairs connected by the supercharges. In the case where the charges do not annihilate the ground state, its energy is non-zero and supersymmetry breaks spontaneously.
\newsection{Encoding the theory on qubits}
To encode the model on a quantum hardware, the operators and degrees of freedom have to be represented in terms of qubits, allowing manipulation by the quantum gates available on the hardware. With this goal in mind, the fermion operators can be encoded by using the well known Jordan-Wigner transformation in Eq.~\ref{eq:Jordan_Wigner}, which describe the fermion state as an occupation state using a single qubit, where the absence of a fermion is denoted by $\vert 0 \rangle$ and its presence by $\vert 1 \rangle $:
\begin{equation} \label{eq:Jordan_Wigner}
\bhat =\frac{1}{2} (X +iY)
\hspace{1.0 cm}
\bhatdag =\frac{1}{2} (X - iY)
\hspace{1.0 cm}
\Rightarrow
\;
\left[\bhatdag, \bhat\right] = -Z
\end{equation}
where $X$, $Y$ and $Z$ are the Pauli quantum gates having the Pauli matrix representation.

The bosonic operators require a more careful consideration, because of their infinite dimensionality, which implies that the operators have to be regularized by using a finite digitization that allows only a finite number of bosonic modes to be present. Even though this is necessary, it is not without consequences. In fact, the explicit truncation of the bosonic operators breaks supersymmetry explicitly, which is recovered by progressively increasing the number of bosonic modes so that the untruncated limit can be extracted. This concretely means that to address the spontaneous supersymmetry breaking or preservation, the model has to be numerically studied by considering systems with a growing number of bosonic modes.

Many different digitizations are possible as indicated in \citep{Macridin:2021uwn}. In this study, the boson number basis is used where the bosonic operators are represented using the position and momentum operators of the harmonic oscillator with a truncation that allows only $\Lambda$ bosonic modes.
Therefore, considering only $\Lambda$ bosonic modes, the $\qhat$ and $\phat$ truncated operators are:
\begin{equation}
\qhat \doteq \frac{1}{\sqrt{2 m}}
\resizebox{0.3\textwidth}{!}
{$
\left(
\begin{array}{ccccc}
0 & \sqrt{1} & 0 &\cdots &0\\
\sqrt{1} & 0 & \sqrt{2} &\cdots &0  \\
0 & \sqrt{2} & \ddots & \ddots &0  \\
0 & 0 & \ddots & 0 & \sqrt{\Lambda -1} \\
0 & 0 & \cdots & \sqrt{\Lambda -1} & 0 \\
\end{array}
\right)
$}
\hspace{0.6 cm}
\phat \doteq i \sqrt{\frac{m}{2}}
\resizebox{0.34\textwidth}{!}
{$
\hspace{0.04 cm}
\left(
\begin{array}{ccccc}
0 & -\sqrt{1} & 0 &\cdots &0\\
\sqrt{1} & 0 & -\sqrt{2} &\cdots &0  \\
0 & \sqrt{2} & \ddots & \ddots &0  \\
0 & 0 & \ddots & 0 & -\sqrt{\Lambda -1} \\
0 & 0 & \cdots & \sqrt{\Lambda -1} & 0 \\
\end{array}
\right)
$}
\end{equation}
Hence, the qubitization of the system consists in one qubit for the fermion, while for the boson, the $\Lambda$ states require $B$ qubits, where $\Lambda =2^B$. This means that the SQM’s Hilbert space composed by $2\Lambda =2^{B+1}$ states can be encoded using $B+1$ qubits, that is an exponential improvement in needed resources.
Therefore, the system can be graphically represented as shown in Fig.~\ref{fig:qubitization_figure}:
\begin{figure}[H]
\centering
\includegraphics[width=0.3\textwidth]{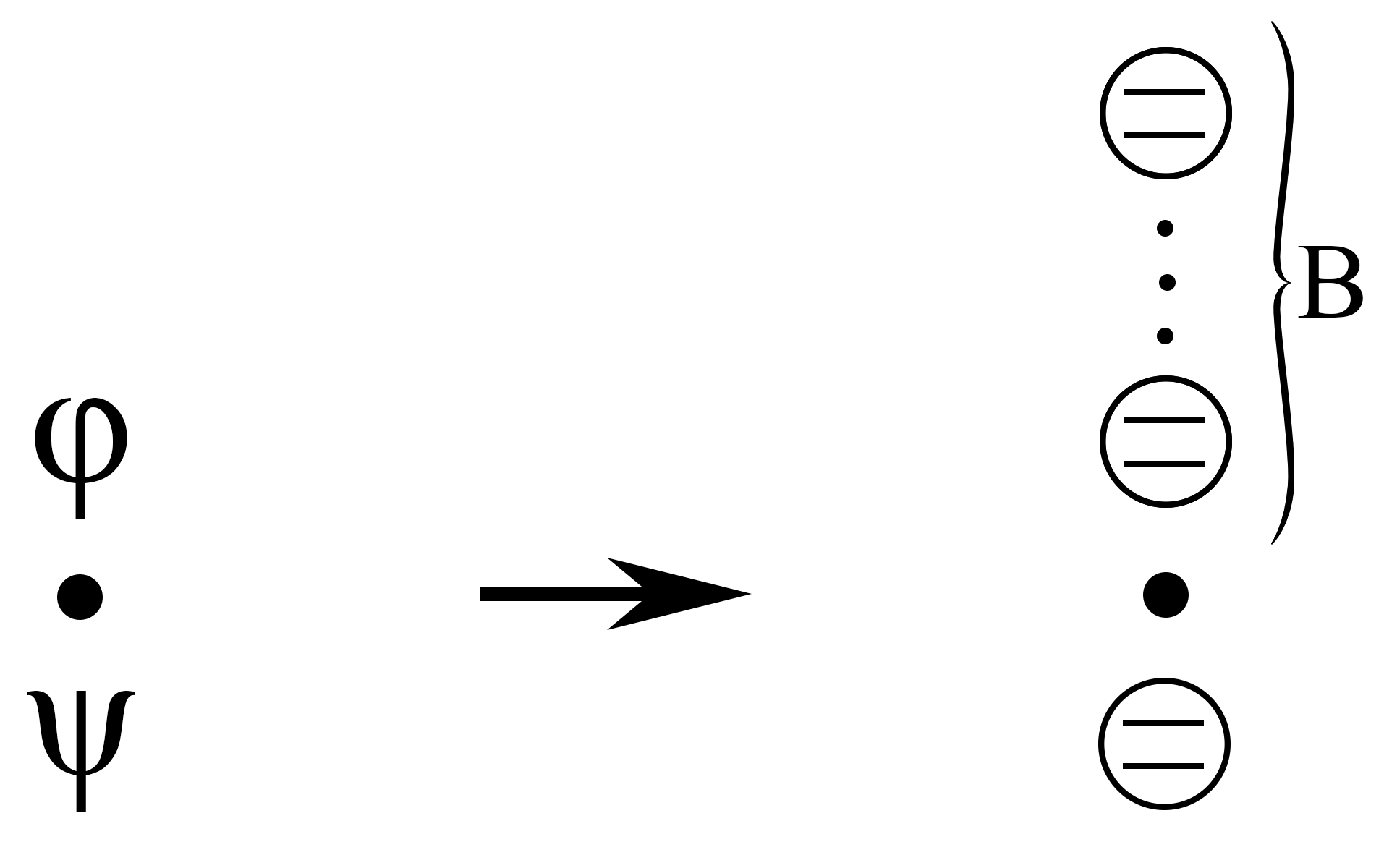}
  \caption{Graphical representation of the qubitization of the model. \textbf{Left:} The untruncated model, where the fermion is represented by $\psi$ while the boson by $\phi$. \textbf{Right:} The model qubitization, where the symbol made by a circle with two lines inside represents a qubit as a two-level system. The fermion needs one qubit while the boson with $\Lambda=2^B$ bosonic modes requires $B$ qubits.}
\label{fig:qubitization_figure}
\end{figure}
To finally encode the model on a quantum hardware, the Hamiltonian needs to be rewritten using the gates available on the hardware, typically the Pauli gates $(I, X, Y, Z)$. This results in the Hamiltonian being expressed as a sum of Pauli strings, formed by the tensor product of as many Pauli operators as the total number of qubits needed for the model. The quantum resources required for each superpotential with varying bosonic modes were obtained using IBM's open-source Qiskit software \citep{qiskit2024} and are detailed in Table~\ref{tab:quantum_resources}.
\begin{table}[h] 
\centering
\begin{tabular}{cccccc}
\hline
\hline
$\Lambda$ & H size & N. qubits & Harmonic Oscillator & Double Well & Anharmonic Oscillator\\ \hline
2 & $4\times 4$ & 2 & 2 & 4 & 2\\
4 & $8\times 8$ & 3 & 4 & 14 &10\\
8 & $16\times 16$ & 4 & 8 & 48 & 34\\
16 & $32\times 32$ & 5 & 16 & 136 & 102\\
32 & $64\times 64$ & 6 & 32 & 352 & 270\\
64 & $128\times 128$ & 7 & 64 & 854 & 670\\
\hline
\hline
\end{tabular}
\caption{Summary of the quantum resources needed to encode the model on an IBM quantum hardware for a growing number of bosonic modes $\Lambda$ and with $(m=g=\mu=1)$ obtained using Qiskit \citep{qiskit2024}. For each superpotential the number of Pauli strings is listed in the respective column.}
\label{tab:quantum_resources} 
\end{table}

\newsection{The variational quantum eigensolver and boxplot visual representation}\label{sec:vqe_boxplot}
The main goal of this study is to investigate on a quantum computer the supersymmetry breaking or preservation of the model by calculating its ground state energy. The variational quantum eigensolver (VQE) \cite{Peruzzo:2013bzg} is employed as our main approach, although other noise-resilient methods such as the quantum imaginary time evolution (QITE) \cite{Motta:2019yya} and the quantum alternating operator ansatz \citep{Farhi:2014ych, Hadfield:2017yqz, Maiti:2024jwk} could also be considered. The Qiskit-based VQE code used for the $0+1$ SQM is provided in Ref.~\citep{0p1SQMcode}.

\begin{figure}[H]
\centering
\includegraphics[width=0.5\textwidth]{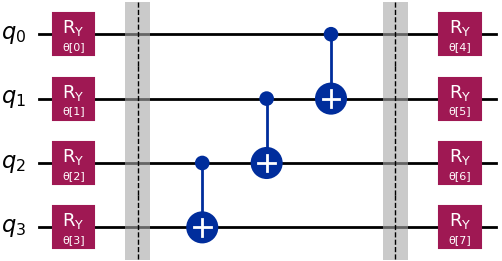}
\caption{The circuit representation of the ‘reverse-linear’ \texttt{Realamplitude} ansatz for the case of a system requiring 4 qubits. The ansatz is made by combining an initial and final layer of rotation gates around the y-axis, individually represented by a square. In the middle there is a layer of entangling gates, CNOTs represented by an elongated cross-like symbol, in reversed order, starting from the last two up to the first two.}
\label{fig:reverse_liear}
\end{figure}

The VQE is a heuristic hybrid quantum-classical algorithm that implements the variational method to find a system's ground state. The algorithm requires that the user proposes an initial parametric form for the state. This state called ansatz is constructed on the quantum hardware in the form of a quantum circuit made by combining a set of single qubit rotations and entangling gates. 
A possible example of a general-purpose ansatz is the one used in this study called \texttt{Realamplitude}, whose circuit representation for a 4-qubit circuit is presented in Fig.~\ref{fig:reverse_liear}. For this ansatz, the number of parameters scales linearly with the number of qubits, requiring $2N_q$ rotation angles and  $N_q-1$ CNOTs.
The algorithm uses the ansatz to measure on the quantum hardware each Pauli string of the Hamiltonian. These measurements are sent to a classical computer, which first calculates the total energy. Then the classical optimizer, in our case the gradient-free Constrained Optimization BY Linear Approximation (COBYLA) algorithm \citep{Powell1994}, tries to minimize the energy by proposing new angles for the ansatz. This cycle can be repeated until a convergence goal is reached.

As a heuristic variational method, VQE does not guarantee finding the true ground state, requiring multiple trials. Therefore, to effectively benchmark the VQE algorithm with a specific ansatz and optimizer, it is essential to control the entire distribution of energy measurements, particularly when dealing with noise or convergence issues. Histograms have traditionally served this purpose, but they can be cumbersome for comparing different data sets.

\begin{figure}[h]
\centering
\includegraphics[width=0.6\textwidth]{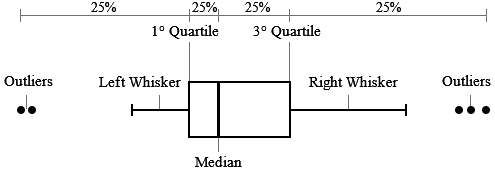}
\caption{Graphical representation of a boxplot with labels for its main components.}
\label{fig:Box_plot_description}
\end{figure}

In this regard, a more concise way is offered by the box and whiskers plot, often abbreviated to boxplots shown in Fig.~\ref{fig:Box_plot_description}, which is a graphical representation of the data where the full distribution of measurements is shown divided into 4 quartiles, each one containing $25\%$ of measurements.
The box contains $50\%$ of the measurements, with the median indicated by a horizontal line separating $25\%$ of measurements to its left from the other $25\%$ to its right.  The lines extending from the box, resembling error bars, are referred to as whiskers. They extend outside the opposite edges of the box up to the last measurement contained in 1.5 times the size of the box. Any measurement that falls outside this range is considered an outlier and individually labelled with a symbol.

\begin{figure}[th]
  \centering
  \includegraphics[width=.49\textwidth]{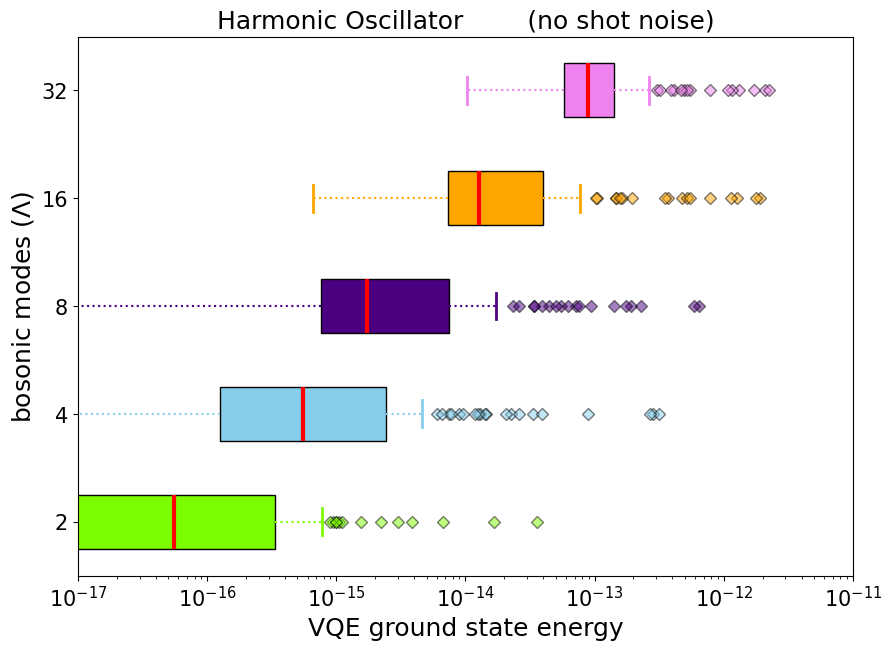}
  \hspace{0.1 cm}
  \includegraphics[width=.49\textwidth]{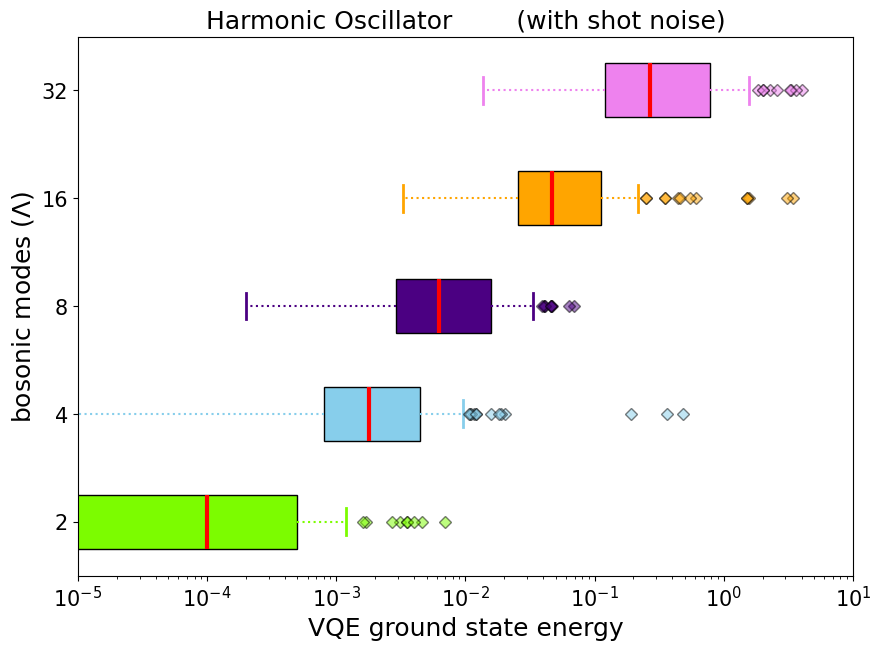}

  \vspace{0.5cm}

  \includegraphics[width=.49\textwidth]{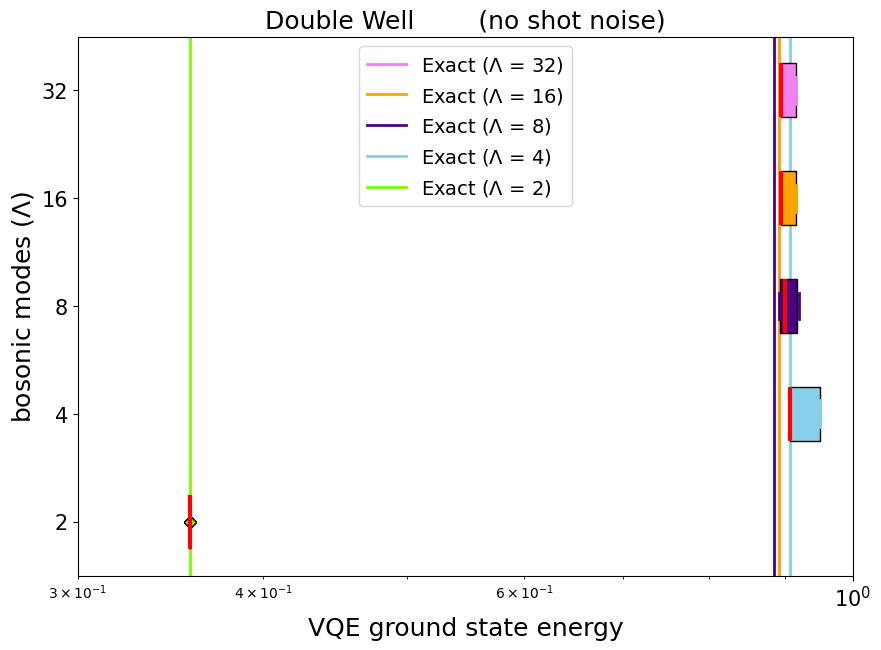}
  \hspace{0.1 cm}
  \includegraphics[width=.49\textwidth]{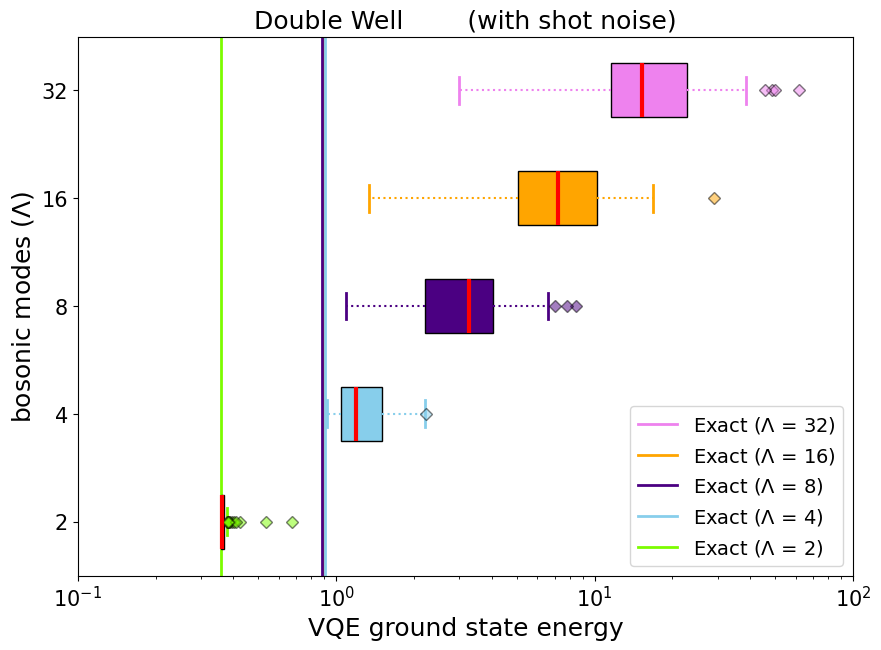}

  \vspace{0.5cm}

  \includegraphics[width=.48\textwidth]{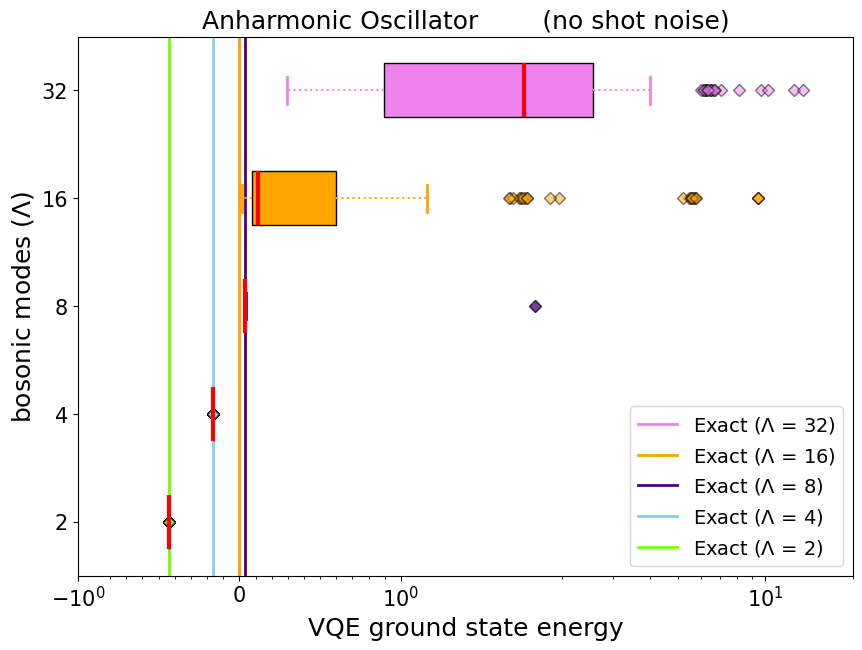}
  \hspace{0.1 cm}
  \includegraphics[width=.49\textwidth]{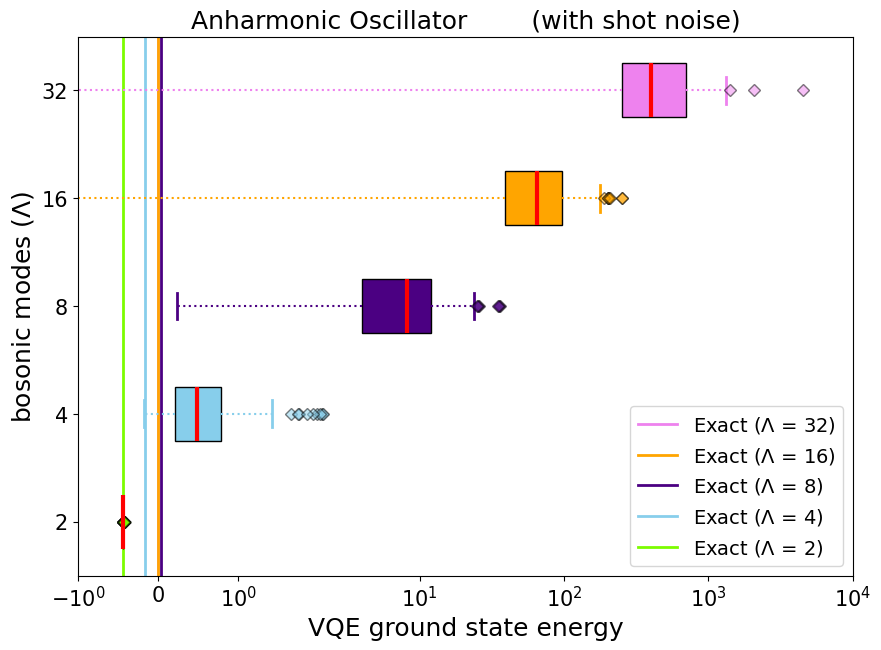}

  \caption{VQE results for the ground state energy of a system with different superpontentials for a growing number of bosonic modes $\Lambda$ presented using boxplots. All the results are obtained using the quantum simulator. \textbf{Left:} The calculation using the exact Qiskit statevector simulator. \textbf{Right:} The calculation using the Qiskit simulator with shot noise, that mimics the intrinsic quantum randomness of a quantum hardware.  Each row contains results for a specific superpotential: Harmonic Oscillator (HO), Double Well (DW) and the Anharmonic Oscillator (AHO)  respectively in order top to bottom. In each plot, the boxplots show the result of 100 VQE runs using COBYLA optimizer with 10\,000 shots for the case with shot noise. The vertical lines indicate the exact classical result obtained through exact diagonalization.}\label{fig:VQE_box_plot}
\end{figure}

For a thorough presentation of boxplot and its use to analyze the periodic trends and properties of chemical elements see \citep{FERREIRA2016209}. Furthermore, an interested reader may find it useful to see how boxplots were used in the field of quantum chemistry by \citep{ Mihalikova:2021oqe}, to present VQE results for the $H_2$ ground state, which deeply inspired us to use them to present our VQE results in Fig.~\ref{fig:VQE_box_plot}.
%
%
%
%
%
\newsection{VQE Results}\label{sec:vqe_results}
Preliminary results for the ground state energy of the SQM model, coming from 100 independent VQE runs with three superpotentials, HO, DW and AHO, and for a growing number of bosonic modes $\Lambda$ are collected in Fig.~\ref{fig:VQE_box_plot}. For each superpotential two set of simulations were conducted. The exact simulation using the Qiskit statevector simulator is shown on the left. On the right, the simulation with shot noise using 10\,000 shot executions for each of the 100 VQE runs.
In both cases, the COBYLA optimizer with the \texttt{RealAmplitude} ansatz was used, typically requiring far fewer than 10,000 optimization iterations, except for statevector simulations of DW and AHO with $\Lambda \geq 16$. The entire procedure was repeated while tightening the optimizer's tolerance from $10^{-1}$ to $10^{-8}$. This approach achieved convergence for most of the VQE statevector runs and for all the VQE runs with shot noise. The exact ground state energy, obtained through exact diagonalization, is displayed by vertical lines, coloured according to the number of bosonic modes used.

The VQE statevector simulator for the HO superpotential case has a distribution of measurements consistent with the exact calculation of zero for all values of $\Lambda$, successfully confirming supersymmetry preservation. In presence of shot noise, the results remain relatively close to zero. However, the median is consistent with a non-zero energy value, which tends to increase with larger $\Lambda$. This illustrates how noise from a fixed number of shots might be misinterpreted as evidence of spontaneous supersymmetry breaking. This issue can be addressed by repeating the calculations with a different number of shots, however this increases computational costs.

In the DW case, the statevector measurements agree with the exact non-zero value for all values of $\Lambda$, successfully confirming spontaneous supersymmetry breaking. Simulations with shot noise behave similarly to HO, with the median moving to steadily larger energy values compared to the exact result and the statevector measurements. Nevertheless, the larger order of magnitude of these values, compared to the HO case, provides evidence of spontaneous supersymmetry breaking even in presence of noise.

Finally, in the AHO case, where the bosonic truncation effect is severe, the ground state energy is non-zero for small $\Lambda$ and exponentially approaches zero as $\Lambda$ is increased, as shown by the vertical lines in Fig.~\ref{fig:VQE_box_plot}. Statevector simulations agree with the exact calculation only up to $\Lambda=8$, making it difficult to confirm supersymmetry preservation. In contrast, simulations with shot noise yield higher and more spread energy measurements, showing discrepancies even at lower values of $\Lambda$.
\newsection{Conclusions and future prospects}\label{sec:conclusions}
In this work, we employed the VQE to investigate spontaneous supersymmetry breaking in SQM with various superpotentials, emphasizing the critical role of shot noise, which has been overlooked in previous studies \citep{Culver:2021rxo}. Preliminary results clearly show that further investigations are necessary to enhance the VQE convergence toward the exact results and to reduce the computational costs of the simulations. This includes reducing the number of independent VQE runs, optimizer iterations, and circuit executions. In Fig.~\ref{fig:VQE_box_plot}, the statevector results reveal that the general purpose \texttt{Realamplitude} ansatz yields reasonable results for systems with a modest number of bosonic modes. However, it does not appear to be sufficient to extract the untruncated behaviour of the theory.

In this respect, a tailored ansatz can be developed by mimicking the entangling structure of the ground state, which for systems with a small number of bosonic modes can be calculated by exact diagonalization. This creates the opportunity to propose a  well shaped ansatz for systems with growing numbers of bosonic modes.

Furthermore, in Fig.~\ref{fig:VQE_box_plot} results with shot noise reveal a tension between the VQE calculation and the exact result, even with a smaller number of bosonic modes, where with no shot noise, there was agreement. This suggests that the COBYLA optimizer struggles to find the minimum in presence of noise. Consequently, future studies will address how to improve the optimizer's performance in presence of shot noise, while also exploring other optimizers, which may offer faster convergence and greater resilience to noise \citep{Pellow_Jarman_2021}. Additionally, efforts will also focus on more efficient digitization techniques to minimize the number of Pauli strings.

The three main directions outlined above are necessary for studying the model on real quantum hardware. Once these steps are fully addressed, it will be crucial to address hardware noise by developing error mitigation techniques for the VQE.

Finally, for a more reliable and noise resilient approach to supersymmetry breaking, the VQE analysis of the ground state should be followed or replaced by the variational quantum deflation (VQD) algorithm \citep{Higgott:2018doo}, which can be used to approximate the ground state and first few excited states. In this way the energy spectrum structure needed to conclude about supersymmetry breaking/preservation can be obtained even in presence of noise. This has already been pointed out in a previous study of the Wess-Zumino model in \citep{Schaich:2024bmg}.

\vspace{8 pt}
\noindent \textsc{Acknowledgments:}
We thank Chris Culver for past collaboration on quantum computing, as well as Nouman Butt, Patrick Draper and Jacopo Settino for helpful discussions about the effects of shot noise in variational quantum algorithms, and John Kerfoot for comments on the draft. This work was supported by UK Research and Innovation Future Leader Fellowship {MR/S015418/1} $\&$ {MR/X015157/1} and STFC grant {ST/T000988/1} $\&$ {ST/X000699/1}.\\

\noindent \textbf{Data Availability Statement:} The data used in this work can be obtained by contacting EM, or reproduced via Ref.~\citep{0p1SQMcode}.

\clearpage
\bibliographystyle{JHEP}
\setlength{\bibsep}{3pt plus 0.3ex}
\bibliography{PoS2024_Bib.bib}

\providecommand{\href}[2]{#2}\begingroup\raggedright\begin{thebibliography}{10}

\bibitem{Kadoh:2016eju}
D.~Kadoh, \emph{{Recent progress in lattice supersymmetry: from lattice gauge
  theory to black holes}},
  \href{https://doi.org/10.22323/1.251.0017}{\emph{PoS} {\bfseries LATTICE2015}
  (2016) 017} [\href{https://arxiv.org/abs/1607.01170}{{\ttfamily
  1607.01170}}].

\bibitem{Bergner:2016sbv}
G.~Bergner and S.~Catterall, \emph{{Supersymmetry on the lattice}},
  \href{https://doi.org/10.1142/S0217751X16430053}{\emph{Int. J. Mod. Phys. A}
  {\bfseries 31} (2016) 1643005}
  [\href{https://arxiv.org/abs/1603.04478}{{\ttfamily 1603.04478}}].

\bibitem{Schaich:2022xgy}
D.~Schaich, \emph{{Lattice studies of supersymmetric gauge theories}},
  \href{https://doi.org/10.1140/epjs/s11734-022-00708-1}{\emph{Eur. Phys. J.
  ST} {\bfseries 232} (2023) 305}
  [\href{https://arxiv.org/abs/2208.03580}{{\ttfamily 2208.03580}}].

\bibitem{Preskill:2018jim}
J.~Preskill, \emph{{Quantum Computing in the NISQ era and beyond}},
  \href{https://doi.org/10.22331/q-2018-08-06-79}{\emph{Quantum} {\bfseries 2}
  (2018) 79} [\href{https://arxiv.org/abs/1801.00862}{{\ttfamily 1801.00862}}].

\bibitem{Peruzzo:2013bzg}
A.~Peruzzo, J.~McClean, P.~Shadbolt, M.-H.~Yung, X.-Q.~Zhou, P.J.~Love et~al.,
  \emph{{A variational eigenvalue solver on a photonic quantum processor}},
  \href{https://doi.org/10.1038/ncomms5213}{\emph{Nature Commun.} {\bfseries 5}
  (2014) 4213} [\href{https://arxiv.org/abs/1304.3061}{{\ttfamily 1304.3061}}].

\bibitem{Culver:2021rxo}
C.~Culver and D.~Schaich, \emph{{Quantum computing for lattice supersymmetry}},
  \href{https://doi.org/10.22323/1.396.0153}{\emph{PoS} {\bfseries LATTICE2021}
  (2022) 153} [\href{https://arxiv.org/abs/2112.07651}{{\ttfamily
  2112.07651}}].

\bibitem{Cai:2022yup}
M.L.~Cai, Y.K.~Wu, Q.X.~Mei, W.D.~Zhao, Y.~Jiang, L.~Yao et~al.,
  \emph{{Observation of supersymmetry and its spontaneous breaking in a trapped
  ion quantum simulator}},
  \href{https://doi.org/10.1038/s41467-022-31058-0}{\emph{Nature Commun.}
  {\bfseries 13} (2022) 3412}
  [\href{https://arxiv.org/abs/2205.14860}{{\ttfamily 2205.14860}}].

\bibitem{Culver:2023iif}
C.~Culver and D.~Schaich, \emph{{Quantum Computing for the
  Wess\textendash{}Zumino Model}},
  \href{https://doi.org/10.22323/1.430.0008}{\emph{PoS} {\bfseries LATTICE2022}
  (2023) 008} [\href{https://arxiv.org/abs/2301.02230}{{\ttfamily
  2301.02230}}].

\bibitem{Schaich:2024bmg}
D.~Schaich and C.~Culver, \emph{{Exploring lattice supersymmetry with
  variational quantum deflation}},
  \href{https://doi.org/10.22323/1.453.0212}{\emph{PoS} {\bfseries LATTICE2024}
  (2024) 212} [\href{https://arxiv.org/abs/2410.11514}{{\ttfamily
  2410.11514}}].

\bibitem{Cooper:1994eh}
F.~Cooper, A.~Khare and U.~Sukhatme, \emph{{Supersymmetry and quantum
  mechanics}}, \href{https://doi.org/10.1016/0370-1573(94)00080-M}{\emph{Phys.
  Rept.} {\bfseries 251} (1995) 267}
  [\href{https://arxiv.org/abs/hep-th/9405029}{{\ttfamily hep-th/9405029}}].

\bibitem{Macridin:2021uwn}
A.~Macridin, A.C.Y.~Li, S.~Mrenna and P.~Spentzouris, \emph{{Bosonic field
  digitization for quantum computers}},
  \href{https://doi.org/10.1103/PhysRevA.105.052405}{\emph{Phys. Rev. A}
  {\bfseries 105} (2022) 052405}
  [\href{https://arxiv.org/abs/2108.10793}{{\ttfamily 2108.10793}}].

\bibitem{qiskit2024}
A.~Javadi-Abhari, M.~Treinish, K.~Krsulich, C.J.~Wood, J.~Lishman, J.~Gacon
  et~al., \emph{Quantum computing with {Q}iskit},  2024.
\newblock 10.48550/arXiv.2405.08810.

\bibitem{Motta:2019yya}
M.~Motta, C.~Sun, A.T.K.~Tan, M.J.O.~Rourke, E.~Ye, A.J.~Minnich et~al.,
  \emph{{Determining eigenstates and thermal states on a quantum computer using
  quantum imaginary time evolution}},
  \href{https://doi.org/10.1038/s41567-019-0704-4}{\emph{Nature Phys.}
  {\bfseries 16} (2019) 205}
  [\href{https://arxiv.org/abs/1901.07653}{{\ttfamily 1901.07653}}].

\bibitem{Farhi:2014ych}
E.~Farhi, J.~Goldstone and S.~Gutmann, \emph{{A Quantum Approximate
  Optimization Algorithm}},  \href{https://arxiv.org/abs/1411.4028}{{\ttfamily
  1411.4028}}.

\bibitem{Hadfield:2017yqz}
S.~Hadfield, Z.~Wang, B.~O'Gorman, E.G.~Rieffel, D.~Venturelli and R.~Biswas,
  \emph{{From the quantum approximate optimization algorithm to a quantum
  alternating operator ansatz.}},
  \href{https://doi.org/10.3390/a12020034}{\emph{Algorithms (Basel)} {\bfseries
  12} (2019) 34} [\href{https://arxiv.org/abs/1709.03489}{{\ttfamily
  1709.03489}}].

\bibitem{Maiti:2024jwk}
S.~Maiti, D.~Banerjee, B.~Chakraborty and E.~Huffman, \emph{{Spontaneous
  symmetry breaking in a $SO(3)$ non-Abelian lattice gauge theory in $2+1$D
  with quantum algorithms}},
  \href{https://arxiv.org/abs/2409.07108}{{\ttfamily 2409.07108}}.

\bibitem{0p1SQMcode}
E.~Mendicelli, ``{VQE codes for $0+1$ supersymmetric quantum mechanics}.''
  \href{https://github.com/emanuele-mendicelli/0p1_Supersymmetric_Quantum_Mechanics}{github.com/emanuele-mendicelli/0p1{\_}Supersymmetric{\_}Quantum{\_}Mechanics},
  2024.

\bibitem{Powell1994}
M.J.D.~Powell, \emph{A direct search optimization method that models the
  objective and constraint functions by linear interpolation},  in
  \emph{Advances in Optimization and Numerical Analysis}, S.~Gomez and
  J.-P.~Hennart, eds., (Dordrecht), pp.~51--67, Springer Netherlands (1994),
  \href{https://doi.org/https://doi.org/10.1007/978-94-015-8330-5_4}{DOI}.

\bibitem{FERREIRA2016209}
J.E.V.~Ferreira, M.T.S.~Pinheiro, W.R.S.~{dos Santos} and R.~da~Silva~Maia,
  \emph{Graphical representation of chemical periodicity of main elements
  through boxplot},
  \href{https://doi.org/https://doi.org/10.1016/j.eq.2016.04.007}{\emph{Educación
  Química} {\bfseries 27} (2016) 209}.

\bibitem{Mihalikova:2021oqe}
I.~Mih\'alikov\'a, M.~Pivoluska, M.~Plesch, M.~Fri\'ak, D.~Nagaj and
  M.~\v{S}ob, \emph{{The Cost of Improving the Precision of the Variational
  Quantum Eigensolver for Quantum Chemistry}},
  \href{https://doi.org/10.3390/nano12020243}{\emph{Nanomaterials} {\bfseries
  12} (2022) 243} [\href{https://arxiv.org/abs/2111.04965}{{\ttfamily
  2111.04965}}].

\bibitem{Pellow_Jarman_2021}
A.~Pellow-Jarman, I.~Sinayskiy, A.~Pillay and F.~Petruccione, \emph{A
  comparison of various classical optimizers for a variational quantum linear
  solver}, \href{https://doi.org/10.1007/s11128-021-03140-x}{\emph{Quantum
  Information Processing} {\bfseries 20} (2021) }.

\bibitem{Higgott:2018doo}
O.~Higgott, D.~Wang and S.~Brierley, \emph{{Variational Quantum Computation of
  Excited States}},
  \href{https://doi.org/10.22331/q-2019-07-01-156}{\emph{Quantum} {\bfseries 3}
  (2019) 156} [\href{https://arxiv.org/abs/1805.08138}{{\ttfamily
  1805.08138}}].

\end{thebibliography}\endgroup

\end{document}